\documentstyle[twoside,fleqn,espcrc2,epsf]{article}

\def\spose#1{\hbox to 0pt{#1\hss}}
\def\ltapprox{\mathrel{\spose{\lower 3pt\hbox{$\mathchar"218$}}
 \raise 2.0pt\hbox{$\mathchar"13C$}}}
\def\gtapprox{\mathrel{\spose{\lower 3pt\hbox{$\mathchar"218$}}
 \raise 2.0pt\hbox{$\mathchar"13E$}}}
\def\inapprox{\mathrel{\spose{\lower 3pt\hbox{$\mathchar"218$}}
 \raise 2.0pt\hbox{$\mathchar"232$}}}

\newcommand{\AmS}{{\protect\the\textfont2
  A\kern-.1667em\lower.5ex\hbox{M}\kern-.125emS}}

\def\lvec#1{\setbox0=\hbox{$#1$}
   \setbox1=\hbox{$\scriptstyle\leftarrow$}
    #1\kern-\wd0\smash{
   \raise\ht0\hbox{$\raise1pt\hbox{$\scriptstyle\leftarrow$}$}}
   \kern-\wd1\kern\wd0}

\newcommand{\chib}{{\bar \chi}}

\newcommand{\zetab}{{\bar \zeta}}

\newcommand{\zetap}{{\zeta^\prime}}

\newcommand{\zetabp}{{\bar \zeta^\prime}}
\newcommand{\vx}{{\vec x}}

\newcommand{\cO}{{\mathcal O}}
\newcommand{\hk}{{\hat k}}

\title{
\vskip -94pt
{\small
 \mbox{} \hfill FSU-SCRI-97C-107\\
 \mbox{} \hfill September 1997\\
}
\vskip  56pt
The Schr\"odinger functional running coupling with staggered
fermions and its application to many flavor QCD\thanks{
To appear in the proceedings of Lattice '97, Edinburgh, Scotland,
July 22--26, 1997.}\thanks{
Work supported by DOE grants DE-FG05-85ER250000 and DE-FG05-96ER40979.}}

\author{Urs~M.~Heller\address{SCRI, The Florida State University,
                            Tallahassee, FL 32306-4130, USA}}

\begin{document}

\begin{abstract}
We discuss the Schr\"odinger functional in lattice QCD with staggered
fermions and relate it, in the classical continuum limit, to the
Schr\"odinger functional regularized with Wilson fermions. We compute the
strong coupling constant defined via the Schr\"odinger functional with
staggered fermions at one loop and show that it agrees with the continuum
running coupling constant in the Schr\"odinger functional formalism. We
compute this running coupling in the ``weak coupling phase'' of many flavor
QCD numerically at several values of the bare coupling and for several
system sizes from $L/a=4$ to 12. The results indicate that the
$\beta$-function for 16 flavors has the opposite sign than for few flavor
QCD, in agreement with a recent claim, and with the perturbative
prediction.

\end{abstract}

\maketitle


\section{THE SCHR\"ODINGER FUNCTIONAL WITH STAGGERED FERMIONS}
\label{stag_Schroed}

The Schr\"odinger functional describes the evolution of a state at
(Euclidean) time $t=0$ to another state at time $t=T$. Using the transfer
matrix it can be written as a path integral with fixed boundary conditions
at time $t=0$ and $T$. For staggered fermions all degrees of freedom can be
fixed at both boundaries \cite{SF_KS}. The Schr\"odinger functional can
thus be represented as the path integral
\begin{eqnarray}
{\mathcal Z}[W,\zeta,\zetab; W^\prime,\zetap,\zetabp] = \int [DU] \\
\int \prod_{\vx}  \prod_{x_4=1}^{T-1} \left[ d\chib(\vx,x_4)
 d\chi(\vx,x_4) \right]  {\rm e}^{-S_G - S_{SF}} . \nonumber 
\end{eqnarray}
Here $W$ and $W^\prime$ represent the boundary values of the gauge fields
and $\zeta$, $\zetab$, $\zetap$ and $\zetabp$ those of the fermion fields.
The pure gauge action and measure are as in \cite{SF_G}. The fermionic part
of the action is the usual staggered action with $\chi$ and $\chib$ set to
zero for $x_4 < 0$ and $x_4 > T$, and to their boundary values at the
boundaries \cite{SF_KS,SF_UMH}.

Note that for staggered fermions the total number of time-slices has to be
even. Therefore the time extent $T$ has to be {\it odd}. In the spatial
direction we take the lattice to be of size $L$ (even!) and impose the
generalized periodic boundary conditions \cite{SF_SW} $\chi(x+L\hk) = {\rm
e}^{i \theta_k} \chi(x)$, $\chib(x+L\hk) = \chib(x) {\rm e}^{-i \theta_k}$.

It can be shown \cite{SF_UMH} that for massless staggered quarks the
classical continuum limit of the Schr\"odinger functional agrees with that
from Wilson fermions \cite{SF_Sint}.

\section{THE SCHR\"ODINGER COUPLING AT ONE LOOP}
\label{coupling_m0}

The ``Schr\"odinger functional coupling constant'' is defined from the
response to a constant chromoelectric background field introduced via the
boundary fields. We use those given in \cite{SF_G,SF_SW}, which depend on a
parameter, $\eta$, and define
\begin{eqnarray}
\label{eq:coupl_def}
\frac{k}{\bar g^2} &\!\!\!\!\!=&\!\!\!\!\! -
 \frac{\partial}{\partial \eta} \log {\mathcal Z} \Big|_{\eta=0}, \\
 k &\!\!\!\!\!=&\!\!\!\!\! 12 \left( \frac{L}{a} \right)^2
 \left[ \sin \left( \frac{2 \pi a^2}{3LT}
 \right) + \sin \left( \frac{\pi a^2}{3LT} \right) \right] , \nonumber 
\end{eqnarray}
with $T=L$ such that $\bar g^2$ depends only on one scale, $\bar g^2 =
\bar g^2(L)$. The normalization $k$ has been chosen such that $\bar g$
equals the bare coupling at tree--level without any cutoff effects.

The one--loop contribution from the staggered fermions to the coupling
constant, $\bar g^2 = g_0^2 + (p_{1,0} + n_f p_{1,1}) g_0^4 + \cO(g_0^6)$,
comes from the derivative of the fermion fluctuation determinant
\begin{equation}
p_{1,1} = \frac{1}{k n_f} \frac{\partial}{\partial \eta} \log \det
 M \Big|_{\eta=0} .
\label{eq:ferm_1_lp}
\end{equation}
Here $n_f=4$ for one flavor of staggered fermions, since they correspond to
four flavors of continuum fermions. The fermion boundary fields are set to
zero, and $M$ is the fermion matrix.

To be precise, the coupling is defined in \cite{SF_G,SF_SW} as
eq.~(\ref{eq:coupl_def}) with $T=L$ so that it depends on a single scale,
$L$. Unfortunately this is not possible for staggered fermions, since, as
we have seen, $L/a$ must be even but $T/a$ must be odd. Instead, we average
the couplings obtained with $T=L+a$ and $T=L-a$ to avoid an additional
$\cO(a)$ effect.

We have evaluated \cite{SF_UMH} $p^{(\pm)}_{1,1}$ of
eq.~(\ref{eq:ferm_1_lp}) for $L/a$ ranging from 4 to 64, in steps of 2.
Here the superscript $\pm$ stands for the choices $T=L \pm a$. One can then
extract the first few coefficients of the expected asymptotic form
\begin{eqnarray}
\label{eq:p11_asym}
p_{1,1}(L/a) &\!\!\!\!\!=&\!\!\!\!\! r_0 + s_0 \log(L/a) \\
 &\!\!\!\!\!+&\!\!\!\!\! (r_1 + s_1 \log(L/a) ) (a/L) +
 \cdots . \nonumber
\end{eqnarray}

$s_0$ should just be $2b_{0,1} = -1/(12\pi^2)$, the fermionic contribution,
per flavor, to the $\beta$-function \cite{SF_SW}, and thus absorbed by
renormalization. We indeed found this result. $s_1$ was found to be
compatible with zero. $r_{0,1}$ are listed in Table~\ref{tab:r0_m0}. $r_0$
obtained from the choices $T=L \pm a$ always agreed to the accuracy given.

\begin{table}
  \begin{center}
    \caption{The first two `non-log' terms in the expansion
             eq.~(\protect\ref{eq:p11_asym}) of $p_{1,1}$. $r^{(av)}_1$
             is the average of the choices $T=L \pm a$.}
    \label{tab:r0_m0}
    \tabcolsep 4pt
    \vspace{2mm}
    \begin{tabular}{|c|c|c|} \hline
     $\theta$ & $r_0$ & $r^{(av)}_1$ \\ \hline
     $0$     & $-0.004416(1)~~$ & $0.00947(1)~$ \\
     $\pi/5$ & $-0.00579695(2)$ & $0.009477(5)$ \\
     $1.0$   & $-0.0068642(1)~$ & $0.00947(1)~$ \\
    \hline
    \end{tabular}
  \end{center}
  \vskip -8mm
\end{table}

$r_0$ of eq.~(\ref{eq:p11_asym}) is the finite part of the fermionic
contribution to the one-loop relation between bare lattice and running
Schr\"odinger functional coupling. When converted to a relation between the
$\overline{\rm MS}$ and the Schr\"odinger functional coupling, we found
\cite{SF_UMH} good agreement with the results of Sint and Sommer from
Wilson fermions \cite{SF_SW}. This confirms that our Schr\"odinger
functional for massless staggered fermions is a correct regularization of
the continuum Schr\"odinger functional with massless fermions.

\subsection{Lattice artefacts}

The vanishing of $s_1$ indicates the absence of bulk $\cO(a)$ artefacts.
The non-vanishing of $r_1^{(av)}$ reflects the presence of boundary $\cO(a)$
effects. These can be absorbed into a pure gauge boundary counterterm
\cite{SF_UMH}.

\begin{figure}
  \vspace{2.5in}
  \includegraphics{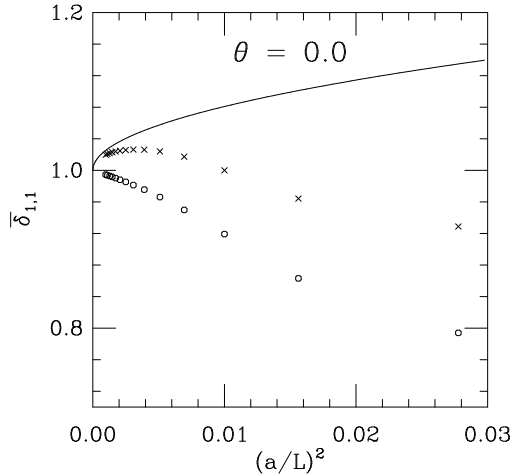}
\vskip -0.3cm
  \caption{The ratio $\overline \delta_{1,1}(a/L)$
           (\protect\ref{eq:delta_11}), for $\theta=0$. Crosses show the
           result before cancellation of the $\cO(a)$ part by the pure
           gauge boundary counterterm, and octagons the result after the
           cancellation. The line shows the $\cO(a)$ part of the lattice
           artefact that is cancelled by the counterterm.}
  \label{fig:sb_0_0}
\vskip -0.3cm
\end{figure}

\begin{figure}
  \vspace{2.5in}
  \includegraphics{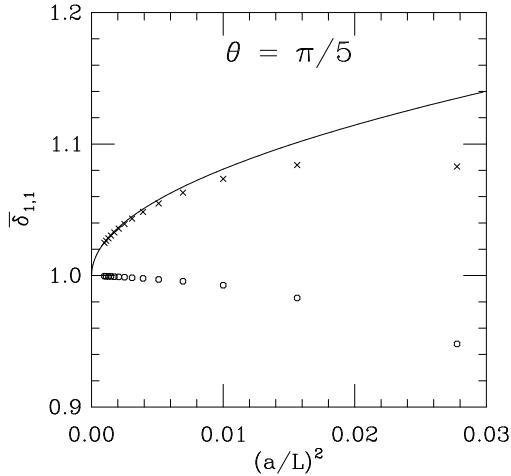}
\vskip -0.3cm
  \caption{Same as Fig.~\protect\ref{fig:sb_0_0}, but for $\theta=\pi/5$.}
  \label{fig:sb_pi5_0}
\vskip -0.3cm
\end{figure}

But higher order lattice artefacts, both from the bulk and the boundary,
remain. We can study them for the fermionic contribution to the step
scaling function \cite{SF_G,SF_SW} with scale factor 2. We compare the
fermion contribution on the lattice (per continuum flavor) with its
continuum limit:
\begin{equation}
\overline \delta_{1,1}(a/L) = \frac{p^{(av)}_{1,1}(2L/a) -
 p^{(av)}_{1,1}(L/a)}{2b_{0,1} \log2} .
\label{eq:delta_11}
\end{equation}
$\overline \delta_{1,1}(a/L)$ is shown in Figs.~\ref{fig:sb_0_0} and
\ref{fig:sb_pi5_0} for $\theta=0$ and $\pi/5$, respectively. Its deviation
from 1 at finite $a/L$ is a lattice artefact.  As can be seen from the two
cases, the higher order lattice artefacts can depend very sensitively on
the observable considered, here the step $\beta$-function for different
values of the spatial boundary conditions.

\section{APPLICATION TO 16 FLAVOR QCD}
\label{QCD_16f}

For $n_f$ between 9 and 16 the perturbative two--loop $\beta$-function has
a second zero, that for $n_f=16$ occurs at a coupling, $g^2 \approx 0.5$,
where the perturbative prediction might be trusted. In a recent study of
$n_f=16$ flavor lattice QCD with staggered fermions we found a bulk first
order transition from a confined, chirally broken phase at strong coupling
to a phase that appeared deconfined and chirally symmetric \cite{DHKO}.
Measurement of masses from correlation functions with pion, rho and nucleon
quantum numbers indicated that the $\beta$-function in this weak coupling
phase might have the opposite sign than few flavor QCD, as perturbation
theory predicts. However, in a deconfined phase it is not so clear, what is
meant by ``hadron masses'' and how to relate their behavior to the
$\beta$-function, which is defined as the change of the coupling under a
change of scale.

The Schr\"odinger functional formalism gives a direct definition of a
running coupling as a function of scale, given by the size of the system.
We have measured this coupling at several values of the bare coupling  in
the ``weak coupling phase'', and for several system sizes, in numerical
simulations. These simulations have been done at {\it zero} quark mass,
which is possible with the fixed boundary conditions in the Schr\"odinger
functional. This avoids having to deal with the anomalous dimension of the
quark mass. The results, with $1/\bar g^2$ --- the observable in the
simulations (see (\ref{eq:coupl_def})) --- averaged over the choices $T = L
\pm 1$, are listed in Table~\ref{tab:g2_nf16}.

\begin{table}
  \begin{center}
    \caption{The Schr\"odinger functional coupling, $\bar g^2(L)$, for
             $n_f=16$, with $1/\bar g^2$ averaged over the choices
             $T=L \pm 1$, for various bare couplings and system sizes.}
    \label{tab:g2_nf16}
    \tabcolsep 4pt
    \vspace{2mm}
    \begin{tabular}{|c|c|c|c|c|} \hline
     $\beta$ & $L=4$ & $L=6$ & $L=8$ & $L=12$ \\ \hline
     4.5 & 5.31(15) & 3.50(16) & 3.22(14) & 2.87(12) \\
     4.6 & 4.41(16) & 2.97(5)~ & 2.77(9)~ &          \\
     4.7 & 3.92(11) & 2.78(4)~ &          &          \\
     4.8 & 3.30(6)~ & 2.67(3)~ & 2.37(3)~ &          \\
     4.9 & 3.07(7)~ & 2.42(3)~ &          &          \\
     5.0 & 2.91(7)~ & 2.28(3)~ &          &          \\
    \hline
    \end{tabular}
  \end{center}
  \vskip -8mm
\end{table}

We see, from Table~\ref{tab:g2_nf16}, that at fixed bare coupling, $\beta$,
the running coupling {\it decreases} with increasing system size. This is
just the opposite behavior than that exhibited by few flavor QCD: the
$\beta$-function of the renormalized coupling has the opposite sign. On the
other hand, demanding that $\bar g^2(L/2,\beta^\prime) = \bar
g^2(L,\beta)$, {\it i.e.} that the lattice spacing at bare coupling
$\beta^\prime$ is a factor of two larger than at $\beta$, we see that for
$n_f=16$ $\beta^\prime$ is larger than $\beta$, again, opposite the
behavior of few flavor QCD. Hence the $\beta$-function of the bare coupling
has the opposite sign. These findings are in agreement with the prediction
from perturbation theory, and confirm the results of \cite{DHKO}. 




\end{document}